\documentstyle{article}
\font\tenrm=cmr10

\title{Exact Results on \(e^+ e^- \rightarrow e^+ e^- + 2\gamma\)
       at SLC/LEP Energies\thanks{ Work
       supported in part by the U.S. DOE contracts DE-FG05-91ER40627 and
       DE-AC03-76ER00515, and by the Polish Ministry of Education grants
       KBN 203809101 and 223729102.}}

\author{
\large S. Jadach\\
\em Institute of Nuclear Physics\\
\em ul. Kawiory 26a, Cracow, Poland
\and
\large B.F.L. Ward\\
\em Department of Physics and Astronomy\\
\em University of Tennessee, Knoxville, TN 37996-1200, USA\\
\em and\\
\em SLAC, P.O. Box 4349\\
\em Stanford University, Stanford, CA 94309, USA
\and and \and
\large S.A. Yost\\
\em Department of Physics and Astronomy\\
\em University of Tennessee, Knoxville, TN 37996-1200, USA}

\date{June, 1992\\ UTHEP-92-0601\\ hep-ph/9211252\\
      Published: \sl Phys.\ Rev.\ \bf D47 \rm(1993) 2682}

\begin{document}

\maketitle
\begin{abstract}
   We use the spinor methods of the CALKUL collaboration, as
   realized by Xu, Zhang and Chang, to calculate the differential cross
   section for \(e^+ e^- \rightarrow e^+ e^- + 2\gamma\) for
   c.m.s.\ energies in the SLC/LEP regime.
   An explicit complete formula for the respective cross section is
   obtained. The leading log approximation is used to check the formula.
   Applications of the formula to high precision luminosity
   calculations at SLC/LEP are discussed.
\end{abstract}
\newpage

\section{Introduction}  \label{INTRODUCTION}

Currently, an unprecedented level of precision has been reached in both
theory and experiment on the cross section $\sigma_\l$ of the basic
luminosity process \(e^+ e^- \rightarrow e^+ e^- + n\gamma\) for SLC/LEP,
and this precision level
has made possible the strongest tests to date of the $SU_{2L}\times U_1$
 theory in $Z^0$ physics\cite{Z0_tests}. The experimental precision is
currently published\cite{exp_prec} as \(.6\%\) whereas the theoretical
precision as calculated by Jadach {\em et al.}\cite{theor_prec} is currently
published
%%##as \(.25\%\) and is based on the pioneering YFS Monte Carlo event
generator as \(.25\%\) and is based on the YFS Monte Carlo event generator
BHLUMI.\footnote{BHLUMI is available from the Computer Physics
Communications program library.} For a detailed illustration of how the
error in the SLC/LEP luminosity enters the various standard model
parameter measurements, see the paper by F. Dydak in Ref.\ \cite{exp_prec}.
Thus, \(1\%\) checks of the standard model in $Z^0$ physics are now
in progress.

In the near term, the experimental error on $\sigma_\l$ at LEP is
expected to improve to the \(.15\%\) or better regime due to imminent
hardware improvements \cite{improvements}. Accordingly, it is important to
improve the theoretical precision on $\sigma_\l$ in
Refs.\ \cite{theor_prec}
to the \(.05\%\) regime in the same near term, so that we will get our first
glimpse of the \(.2\%\) tests of the $SU_{2L}\times U_1$ theory in $Z^0$
physics. Accordingly, the contributions to the error \(\Delta\sigma_{\cal L}\)
in Table 3 of the third paper in Ref.\ \cite{theor_prec} which violate our
\(.05\%\) requirement for the total error must be computed with an
appropriately improved precision. The largest contribution to
\(\Delta\sigma_\l\) in that table is due to the missing second order
bremsstrahlung contribution, which is itself \(.15\%\). Hence, in this paper,
we compute the exact expression for the differential cross section for the
respective process \(e^+ e^- \rightarrow e^+ e^- + 2\gamma\)
in the SLC/LEP energy regime, with the
understanding that the \(e^+ e^-\) scattering angles
$\theta_{{\overline e} e}$ are always much larger than \(2m_e/\sqrt{s}\).

Specifically, we shall employ methods originally pioneered by the CALKUL
collaboration in Refs.\ \cite{CALKUL1,CALKUL2}. We use these methods
in the manner of Xu, Zhang and Chang in Refs.\ \cite{chinese}. Indeed,
the CALKUL collaboration computed the process
\( e^+ e^- \rightarrow \gamma \rightarrow \mu^+ \mu^- + 2\gamma \)
in Ref.\ \cite{CALKUL2} using their original formulation of the spinor
product method. In Refs.\ \cite{s-channel}, Jadach {\em et al.}\ extended
the CALKUL computation to include $Z^0$ exchange in
\( e^+ e^- \rightarrow \mu^+ \mu^- + 2\gamma \),
thereby arriving at a formula of direct importance to high precision
$Z^0$ physics for the
\( e^+ e^- \rightarrow Z^0 \rightarrow \mu^+ \mu^- + n\gamma \)
process. Entirely similar results, using the methods derived from those
of Xu {\em et al.}, were obtained essentially simultaneously by Kleiss and
van der Marck in Ref.\ \cite{kleiss}. Hence, what we present in this
paper is the natural extension of the work of
Refs.\ \cite{CALKUL2,chinese,kleiss} to the double radiative Bhabha
scattering process \(e^+ e^- \rightarrow e^+ e^- + 2\gamma\).

We should emphasize that the process which we compute has applications
in the $Z^0$ regime beyond just the luminosity. For, recently, there
has been interest in wide angle events with two hard photons\cite{wide_angle}
at \( \sqrt{s} \sim M_{Z^0} \) in
\( e^+ e^- \rightarrow f{\overline f} + 2\gamma \),
with \( f = e, \mu,\) {\em etc}. Here, we present results which are directly
relevant to these events.

Our work is organized as follows. In Sec.\ 2, we review the relevant
spinor product notation and set our kinematic and notation conventions.
In Sec.\ 3, we analyze \(e^+ e^- \rightarrow e^+ e^- + 2\gamma\) and
present a formula for the complete
differential cross section. In Sec.\ 4, we compare our results with
known leading logarithm approximations and thereby determine the size of
the next-to-leading term in the second order bremsstrahlung correction to
the luminosity process at SLC/LEP. Sec.\ 5 contains some concluding remarks.

\section{Preliminaries}
\label{SPINORS}

In order that our analysis is self-contained, we begin it in this
section by stating our notations and conventions. This will also
facilitate comparisons of our work with related efforts in the
literature.

We first note that our metric will be that of Bjorken and
Drell\cite{bjdrell}. Our notation for Dirac gamma matrices will follow
that of Ref.\ \cite{s-channel} in the so-called chiral basis.
Similarly, our conventions for the left and right handed couplings
of the $Z^0$ to the electron are those of Ref.\ \cite{s-channel}
so that
\begin{equation} g_L = e\; \cot 2\theta_W , \qquad
%g_R = -e\; { \sin^2\theta_W\over 2\sin2\theta_W} or more simply:
g_R = -{e      }\tan\theta_W,
\label{coup}\end{equation}
where $e$ is the electric charge of the electron, so that it is negative.
The rest mass of the $Z^0$ is denoted by $M_Z$,which we
take\cite{Z0_tests} as 91.187 GeV. The $Z^0$ width will be denoted by
$\Gamma_Z$ and will be taken\cite{Z0_tests} as 2.492 GeV.

Our conventions for the metric, Dirac $\gamma$ matrices and $Z^0$
charges we then complete with our spinor notation from
Ref.\ \cite{chinese}. Specifically, a massless spinor of four-momentum
$p$ and helicity $\lambda$ is denoted by
\begin{equation} \begin{array}{l}
  \left|{p,\lambda}\right\rangle \equiv u_{\lambda}(p) = v_{-\lambda}(p), \\
  \left\langle {p,\lambda}\right| \equiv {\overline u}_{\lambda}(p) =
       {\overline v}_{-\lambda}(p),
\end{array} \label{spr} \end{equation}
with the normalization
\begin{equation}  \left\langle {p,\lambda|\gamma^{\mu}|p,\lambda}
\right\rangle = 2p^{\mu} .
\label{norm}\end{equation}
These spinors have a number of useful properties, a representative
summary of which may be found in Ref.\ \cite{chinese}.
Here, we finalize our notational discussion of these objects by
introducing the basic unit in which our amplitudes for our Bhabha
scattering process will be expressed,namely, the spinor product.
For two massless four-vectors $p$, $q$, we define the spinor product as
\begin{equation} \left\langle {p,q}\right\rangle_{\lambda} =
\left\langle {p,-\lambda|q,\lambda}\right\rangle .
\label{sprod}\end{equation}

It is sometimes convenient to introduce the triple product
\begin{equation}  \left\langle {p,k,q}\right\rangle_{\lambda} =
\left\langle {p,\lambda}
\right|
\not{\hbox{\kern-2.3pt $k$}} \left|{q,\lambda}\right\rangle,
\label{triprd} \end{equation}
where k is an arbitrary four-vector which need not be massless.
This product is obviously linear in $k$, and when $k$ is massless,
it factorizes into a pair of spinor products:
\begin{equation} \left\langle {p,k,q}\right\rangle_{\lambda} =
\left\langle {p,k}\right\rangle_{-\lambda}
\left\langle {k,q}\right\rangle_{\lambda}
\hbox{\qquad if \quad $k^2 = 0$.}
\label{tpfact} \end{equation}

Using the chiral basis for Dirac matrices in Ref.\ \cite{s-channel},
the light-cone notation $p^{\pm}=p^0\pm p^3$ with the 3-axis directed
along the incoming positron beam direction, and the complex perpendicular
variable $p^{\perp}= p^1 + i p^2$, it can be shown that
\begin{equation} \left|{p,+}\right\rangle = {1\over \sqrt{p^+}}
               \pmatrix{p^+\cr p^{\perp}\cr 0 \cr 0}, \qquad
   \left|{p,-}\right\rangle = {1\over \sqrt{p^+}}\pmatrix{0\cr 0\cr
-p^{\perp *}\cr p^+},
\label{explicit_a} \end{equation}
\begin{equation} \left\langle {p,+}\right| = {-1\over \sqrt{p^+}}
\left(0, 0, p^+, p^{\perp *}\right),
   \qquad \left\langle {p,-}\right| = {1\over \sqrt{p^+}}
\left(p^{\perp}, -p^+, 0, 0\right)
\label{explicit_b}\end{equation}
via standard Dirac equation manipulations. Under the crossing
transformation,
we need the analytic continuation of the square roots in
(\ref{explicit_a}--\ref{explicit_b})
to negative values of $p^+$. We may use
\begin{equation}
\left|{-p,\lambda}\right\rangle = i\left|{p,\lambda}\right\rangle, \qquad
\left\langle {-p,\lambda}\right| = i\left\langle {p,\lambda}\right|.
\label{contin} \end{equation}

Evidently, we may now express our basic spinor products as
explicit functions of their four-vector arguments. We get
\begin{equation} \begin{array}{l}
\left\langle {p,q}\right\rangle_+= {\displaystyle
{1\over \sqrt{p^+}\sqrt{q^+}}}
    \left(p^{\perp}q^+ - p^+q^{\perp}\right),\\ % DON'T COMBINE THE SQ. ROOTS
\left\langle {p,q}\right\rangle_- = \mathop{\rm sgn} p^+
\ \mathop{\rm sgn} q^+ \left\langle {q,p}\right\rangle_+^*,
\end{array} \label{comp2} \end{equation}
where the signs in the second expression are needed to account for the
analytic continuation (\ref{contin}). For the triple product, we get
\begin{equation} \begin{array}{c}
\left\langle {p,k,q}\right\rangle_+ = {\displaystyle
{1\over \sqrt{p^+}\sqrt{q^+}}}
      \left(p^+k^-q^+-p^+k^{\perp *} q^{\perp}-p^{\perp *}k^{\perp}q^+
      +p^{\perp *}k^+q^{\perp}\right), \\     % DON'T COMBINE THE SQ. ROOTS
\left\langle {p,k,q}\right\rangle_- = \left\langle {q,k,p}\right\rangle_+ .
\end{array} \label{comp3} \end{equation}

Finally, we note our convention for the photon polarizations, which
we take from Ref.\ \cite{chinese}. For a photon of helicity
$\lambda$ and four-momentum $k$, we define the polarization four-vector
\begin{equation}\epsilon^{\sigma}(k,l,\lambda) = \lambda {\left\langle
{l,-\lambda|\gamma^{\sigma}|
  k,-\lambda}\right\rangle \over \sqrt2
\left\langle {l,-\lambda|k,\lambda}\right\rangle},
\label{pol}\end{equation}
where $l$ is a reference momentum such that $l^2=k^2=0$.

This completes the notation needed for our analysis. We turn now to
computation of the amplitudes of interest to us in the next section.

\section{Exact two-photon Bremsstrahlung amplitudes}
\label{BHABHA}

In this section, we give exact expressions for all \({\cal O}(\alpha^2)\)
terms in the amplitude \(e^+ e^- \rightarrow e^+ e^- + 2\gamma\),
in the limit where the electron mass
is negligable. The spinor conventions described in Sec.\ \ref{SPINORS}
allows these amplitudes to be expressed in a very compact form. The case
of initial state radiation in the $s$ channel has already been obtained
using these methods by Kleiss and van der Marck\cite{kleiss}. The complete
$s$ channel result, with muons in the final state, has also been obtained
by Jadach {\em et al.}\cite{s-channel}, but in a much less compact form.

The following kinematic conventions will be used.
The momenta of the incoming and outgoing electron will be
denoted $p$ and $q$, while the corresponding helicities will
be denoted $\lambda$ and $\mu$, respectively. The positron variables will
be the same, but with a prime. The photon momenta and helicities will be
denoted $k_i$ and $\rho_i$, respectively.

The total amplitude may be written
as a sum ${\cal M} = {\cal M}^s + {\cal M}^t$ of $s$ and $t$ channel
contributions, with
\begin{equation} \begin{array}{l}
{\cal M}^t = {\cal M}_{pp} + {\cal M}_{ee} + {\cal M}_{pe}, \\
{\cal M}^s = {\cal M}_{ii} + {\cal M}_{ff} + {\cal M}_{if}.
\end{array} \label{terms} \end{equation}
Here, the subscripts $p$, $e$, $i$ and $f$ indicate that a photon is
emitted from the positron line, electron line, an initial state line,
and a final state line,
respectively. Helicity conservation plays an important role in
simplifying the individual terms, as it is explained in
Refs.\ \cite{chinese,kleiss}.

We begin with the $t$ channel. Helicity conservation requires
\( \lambda = \mu\), \( \lambda' = \mu'\) in all nonzero amplitudes.
It is convenient to express these amplitudes
in terms of helicity-dependent momentum variables
\begin{equation}
h_i  = \cases{p \cr q \cr}
\hbox{\ if $\rho_i = \pm\lambda$,} \qquad
h'_i = \cases{p' \cr q' \cr}
\hbox{\ if $\rho_i = \pm\lambda'$,}
\label{hdef} \end{equation}
and to define the $t$-variables
\begin{equation}
t = (p - q)^2, \qquad t' = (p' - q')^2, \qquad t_i = (p - q - k_i)^2.
\label{tdef} \end{equation}
When the photon helicities are identical, {\em i.e.}
\(\rho_1 = \rho_2 \equiv \rho\), the three terms in the $t$ channel
amplitude are given by
\begin{eqnarray}
{\cal M}_{pp} &=& 4ie^4 G_{\lambda,-\lambda'}(t)\;
  {\left\langle {h,h'}\right\rangle_\rho^2 \left\langle {p',q'}
\right\rangle_\rho
\left\langle {p,q}\right\rangle_{-\rho} \over
   \left\langle {k_1,q',k_2}\right\rangle_\rho \left\langle
{k_1,p',k_2}\right\rangle_\rho},
\label{equal-t_a}\\
{\cal M}_{ee} &=& 4ie^4 G_{\lambda,-\lambda'}(t')\;
  {\left\langle {h,h'}\right\rangle_\rho^2 \left\langle
  {q,p}\right\rangle_\rho
\left\langle {q',p'}\right\rangle_{-\rho} \over
   \left\langle {k_1,p,k_2}\right\rangle_\rho \left\langle
{k_1,q,k_2}\right\rangle_\rho},
\label{equal-t_b}\\
{\cal M}_{pe} &=& -4ie^4 \left\langle {h,h'}\right\rangle_\rho^2
        \left\{ {t_1 G_{\lambda,-\lambda'}(t_1) \over
   \left\langle {p,k_1,q}\right\rangle_\rho \left\langle {p',k_2,q'}
\right\rangle_\rho}
   + (1 \leftrightarrow 2) \right\} .
\label{equal-t_c}
\end{eqnarray}
The propagator factor for photon and $Z^0$ exchange is defined by
\begin{equation}
G_{\lambda,\mu}(z) = {1\over z} + {\left[(1-\lambda)
   - 4\sin^2\theta_W\right]
   \left[(1-\mu) - 4\sin^2\theta_W\right] \over 4\sin^2 2\theta_W
   \left[z\left(1 + i\Gamma_Z/M_Z\right) - M_Z^2\right]}.
\label{prop} \end{equation}
The $Z^0$ width $\Gamma_Z$ is to be omitted in the $t$ channel,
but is present in the $s$ channel expressions below.

The same amplitudes for opposite photon helicities are given by the more
complicated expressions
\begin{eqnarray}
{\cal M}_{pp} &=& {4ie^4\delta_{\rho_j,\lambda'}\ G_{\lambda,-\lambda'}(t)
   \over \left\langle {k_i,q',k_j}\right\rangle_{\lambda'}
\left\langle {k_i,p',k_j}\right\rangle_{\lambda'}} \left\{
   \left\langle {h_i,(q'+k_j),p'}\right\rangle_{\lambda'}
\left\langle {q',(p'-k_i),h_j}\right\rangle_{\lambda'}
   \right.\nonumber\\
   &+& {{\widehat\Delta}}^{'-1} \left\langle {q',k_j}\right\rangle_{-\lambda'}
\left\langle {p',h_j}\right\rangle_{\lambda'}
   \left\langle {h_i,(q' + k_j),k_i}\right\rangle_{\lambda'}
\left\langle {k_i,p',k_j}\right\rangle_{\lambda'} \nonumber\\
   &+& \left.{\Delta}^{'-1} \left\langle {p',k_i}\right\rangle_{\lambda'}
\left\langle {q', h_i}\right\rangle_{-\lambda'}
   \left\langle {k_j,(p' - k_i),h_j}\right\rangle_{\lambda'}
\left\langle {k_i,q',k_j}\right\rangle_{\lambda'}
   \right\},
\label{opp-t_a}\\
{\cal M}_{ee} &=& {4ie^4\delta_{\rho_i,\lambda}\ G_{\lambda,-\lambda'}(t')
   \over \left\langle {k_j,p,k_i}\right\rangle_{\lambda}
\left\langle {k_j,q,k_i}\right\rangle_{\lambda}} \left\{
   \left\langle {h'_j,(q+k_i),p}\right\rangle_{\lambda}
\left\langle {q,(p-k_j),h'_i}\right\rangle_{\lambda}
   \right.\nonumber\\
   &+& {\Delta}^{-1} \left\langle {p,k_j}\right\rangle_{\lambda}
\left\langle {q,h'_j}\right\rangle_{-\lambda}
   \left\langle {k_i,(p - k_j),h'_i}\right\rangle_{\lambda}
\left\langle {k_j,q,k_i}\right\rangle_{\lambda} \nonumber\\
   &+& \left.{{\widehat\Delta}}^{-1} \left\langle
{p,h'_i}\right\rangle_{\lambda}
\left\langle {q, k_i}\right\rangle_{-\lambda}
   \left\langle {h'_j,(q + k_i),k_j}\right\rangle_{\lambda}
\left\langle {k_j,p,k_i}\right\rangle_{\lambda}
   \right\},
\label{opp-t_b}\\
{\cal M}_{pe} &=& 4ie^4\left\{ G_{\lambda,-\lambda'}(t_1)
   {\left\langle {h'_2, (\lambda h_2 + \rho_1 k_1), h_1}
\right\rangle_{\rho_1}^2
   \over \left\langle {p,k_1,q}\right\rangle_{\rho_1} \left\langle
{p',k_2,q'}\right\rangle_{\rho_2}}
   + (1 \leftrightarrow 2) \right\} .
\label{opp-t_c}
\end{eqnarray}
The indicies \((i,j) = (1,2)\) or \((2,1)\) are chosen so that
(\ref{opp-t_a}--\ref{opp-t_b}) are nonzero,
and the denominators are defined by
\begin{equation} \begin{array}{lll}
\Delta = (p - k_1 - k_2)^2, &\qquad& \Delta' = (p' - k_1 - k_2)^2, \\
{\widehat\Delta} = (q + k_1 + k_2)^2, &\qquad& {\widehat\Delta}' =
 (q' + k_1 + k_2)^2.
\end{array} \label{denoms} \end{equation}

The $s$ channel results are analogous. In this case, helicity conservation
requires \(\lambda' = -\lambda\) and \(\mu' = -\mu\) for nonzero amplitudes.
We define the helicity-dependent momentum variables
\begin{equation}
l_i  = \cases{p \cr p' \cr}
\hbox{\ if $\rho_i = \pm\lambda$,} \qquad
{\widehat l}_i = \cases{q \cr q' \cr}
\hbox{\ if $\rho_i = \mp\mu$,}
\label{ldef} \end{equation}
and the $s$ variables
\begin{equation}
s = (p + p')^2, \qquad {\widehat s} = (q + q')^2, \qquad s_i
= (p + p' - k_i)^2.
\label{sdef} \end{equation}
The three terms in the $s$ channel
amplitude are given by
\begin{eqnarray}
{\cal M}_{ff} &=& 4ie^4 \lambda\mu G_{\lambda,\mu}(s)\;
  {\left\langle {l,{\widehat l}\,}\right\rangle_\rho^2 \left\langle
  {q,q'}\right\rangle_\rho
   \left\langle {p,p'}\right\rangle_{-\rho} \over
   \left\langle {k_1,q,k_2}\right\rangle_\rho \left\langle {k_1,q',k_2}
\right\rangle_\rho},
\label{equal-s_a}\\
{\cal M}_{ii} &=& 4ie^4 \lambda\mu G_{\lambda,\mu}({\widehat s})\;
  {\left\langle {l,{\widehat l}\,}\right\rangle_\rho^2 \left\langle
  {p,p'}\right\rangle_\rho
   \left\langle {q,q'}\right\rangle_{-\rho} \over
   \left\langle {k_1,p,k_2}\right\rangle_\rho \left\langle {k_1,p',k_2}
\right\rangle_\rho},
\label{equal-s_b}\\
{\cal M}_{if} &=& 4ie^4 \lambda\mu \left\langle {l,{\widehat l}\,}
\right\rangle_\rho^2
        \left\{ {s_1 G_{\lambda,\mu}(s_1) \over
   \left\langle {p,k_1,p'}\right\rangle_\rho \left\langle {q,k_2,q'}
\right\rangle_\rho}
   + (1 \leftrightarrow 2) \right\}
\label{equal-s_c}
\end{eqnarray}
for equal photon helicities, and by
\begin{eqnarray}
{\cal M}_{ff} &=& {-4ie^4\delta_{\rho_i,\mu}\ G_{\lambda,\mu}(s)
   \over \left\langle {k_j,q,k_i}\right\rangle_{\mu} \left\langle
{k_j,q',k_i}\right\rangle_{\mu}} \left\{
   \left\langle {q,(q'+k_j),l_i}\right\rangle_{\mu}
\left\langle {l_j,(q+k_i),q'}\right\rangle_{\mu}
   \right.\nonumber\\
   &+& {{\widehat\Delta}}^{'-1} \left\langle {q',k_j}\right\rangle_{\mu}
\left\langle {q,l_j}\right\rangle_{-\mu}
   \left\langle {k_i,(q' + k_j),l_i}\right\rangle_{\mu} \left\langle
{k_j,q,k_i}\right
\rangle_{\mu} \nonumber\\
   &+& \left.{{\widehat\Delta}}^{-1} \left\langle {q,k_i}\right\rangle_{-\mu}
\left\langle {q', l_i}\right\rangle_{\mu}
   \left\langle {l_j,(q + k_i),k_j}\right\rangle_{\mu} \left\langle
{k_j,q',k_i}\right\rangle_{\mu}
   \right\},
\label{opp-s_a}\\
{\cal M}_{ii} &=& {-4ie^4\delta_{\rho_i,\lambda}\ G_{\lambda,\mu}({\widehat s})
   \over \left\langle {k_j,p,k_i}\right\rangle_{\lambda}
\left\langle {k_j,p',k_i}\right\rangle_{\lambda}} \left\{
   \left\langle {{\widehat l}_j,(p'-k_i),p}\right\rangle_{\lambda}
\left\langle {p',(p-k_j),{\widehat l}_i}\right\rangle_{\lambda}
   \right.\nonumber\\
   &+& {\Delta}^{-1} \left\langle {p,k_j}\right\rangle_{\lambda}
\left\langle {p',{\widehat l}_j}\right\rangle_{-\lambda}
   \left\langle {k_i,(p - k_j),{\widehat l}_i}\right\rangle_{\lambda}
\left\langle {k_j,p',k_i}\right\rangle_{\lambda} \nonumber\\
   &+& \left.{\Delta}^{'-1} \left\langle {p,{\widehat l}_i}
\right\rangle_{\lambda}
\left\langle {p', k_i}\right\rangle_{-\lambda}
   \left\langle {{\widehat l}_j,(p' - k_i),k_j}\right\rangle_{\lambda}
\left\langle {k_j,p,k_i}\right\rangle_{\lambda}
   \right\},
\label{opp-s_b}\nonumber\\
{\cal M}_{if} &=& 4ie^4\lambda\mu \left\{ G_{\lambda,\mu}(s_1)
   {\left\langle {{\widehat l}_2, (l_2 - k_1), l_1}\right\rangle_{\rho_1}^2
   \over \left\langle {p,k_1,p'}\right\rangle_{\rho_1}
\left\langle {q,k_2,q'}\right\rangle_{\rho_2}}
   + (1 \leftrightarrow 2) \right\}
\label{opp-s_c}
\end{eqnarray}
for opposite photon helicities.

The amplitudes ${\cal M}$ above are related by crossing symmetries.
Interchanging the incoming positron and outgoing electron, so that
\begin{equation}
p' \leftrightarrow -q, \qquad \lambda' \leftrightarrow -\mu,
\label{switch1} \end{equation}
interchanges the $s$ and $t$ channel amplitudes
\begin{equation}
{\cal M}_{pp} \leftrightarrow {\cal M}_{ff}, \qquad
{\cal M}_{ee} \leftrightarrow {\cal M}_{ii}, \qquad
{\cal M}_{pe} \leftrightarrow {\cal M}_{if},
\label{cross1} \end{equation}
while interchanging the incoming electron and outgoing positron, so that
\begin{equation}
p \leftrightarrow -q', \qquad \lambda \leftrightarrow -\mu',
\label{switch2} \end{equation}
gives
\begin{equation}
{\cal M}_{pp} \leftrightarrow {\cal M}_{ii}, \qquad
{\cal M}_{ee} \leftrightarrow {\cal M}_{ff}, \qquad
{\cal M}_{pe} \leftrightarrow {\cal M}_{if}.
\label{cross2} \end{equation}
These expressions are useful in practice, since, together, they allow
all of the amplitudes to be obtained from only four expressions,
for example, (\ref{equal-t_a}), (\ref{equal-t_c}), (\ref{opp-t_a}), and
(\ref{opp-t_c}). The form of the
expression depends only on whether the photon helicities are equal or
opposite, and whether they are emitted from the same or different fermion
lines.

This completes our derivation of the exact second order matrix
element for the process $e^+e^-\rightarrow e^+e^-+2\gamma$. We turn
now to some of its applications. This we do in the next section.

\section{Sample Results} \label{LEADLOG}        % section 4

In this section we illustrate our exact result in the context of
its main purpose, which is to check the non-leading bremsstrahlung
correction for two hard photons in the low angle regime of Bhabha
scattering in our SLC/LEP luminosity calculations in
Ref.\ \cite{theor_prec}. Thus, we will compare our exact results
with the leading log expectations in the low angle Bhabha scattering
luminosity regime.
We begin with a discussion of the differential cross section for
associated with our exact result.

Using entirely standard manipulations\cite{bjdrell}, we get the
following expression for the exact differential cross section
for $e^+e^-\rightarrow e^+ e^-+2\gamma$ in the $Z^0$ resonance region:
\begin{equation}
d\sigma = {1\over 2!} \delta^4(p + p' - q - q' - k_1 - k_2)
    {\left|{\overline{\cal M}}\right|^2 \over (2\pi)^8 s}
    {d^3{\bf q}\, d^3{\bf q'}\, d^3{\bf k_1}\, d^3{\bf k_2}
     \over 2^4 q^0 q^{'0} k_1^0 k_2^0},
\label{dcs}\end{equation}
where the averaged and summed squared matrix element
may be expressed as a sum over helicities
\begin{equation}
\left|{\overline{\cal M}}\right|^2 \equiv {1\over 4}
\sum_{\lambda,\lambda',\mu,\mu',\rho_1,\rho_2 = \pm1}
     \left|{\cal M}(\lambda,\lambda',\mu,\mu',\rho_1,\rho_2)\right|^2,
\label{avgsum} \end{equation}
using the total exact amplitude derived in the previous section.
We will compare the exact cross section (\ref{dcs}) with the
leading log expectations.
This comparison requires a leading log approximation to the exact
cross section, namely
\begin{equation}
d\sigma_{{}_{\scriptstyle LL}} = {1\over 2!}
     \delta^4(p + p' - q - q' - k_1 - k_2)
    {\left|{\overline{\cal M}}\right|^2_{LL} \over (2\pi)^8 s}
    {d^3{\bf q}\, d^3{\bf q'}\, d^3{\bf k_1}\, d^3{\bf k_2}
     \over 2^4 q^0 q^{'0} k_1^0 k_2^0},
\label{lldcs}\end{equation}
where the leading log summed squared matrix element is given
by (\ref{llsum}) below. It is to the derivation of the latter equation
that we now turn.

For our leading log representation of $\left|{\overline{\cal M}}
\right|^2_{LL}$, we follow the development
given by Jadach and Ward in Ref.\ \cite{yfs2}. Specifically, we
generalize the initial state $s$-channel double bremsstrahlung
leading log differential cross section in Ref.\ \cite{yfs2} to include
analogous contributions from the final state and all other channels
containing collinear singularities. This leads to an expression
for the leading log version of the squared matrix element (\ref{avgsum})
given as a sum of six similar terms, labelled by the fermion lines
emitting the photons (initial, final, positron, electron, or mixed):
\begin{equation}
\left|{\overline{\cal M}}\right|^2_{LL} \equiv \left| {\cal L}_{i}
+ {\cal L}_{f} +
{\cal L}_{p}
 + {\cal L}_{e} - {\cal L}_{m} - {\cal L}_{m'} \right|.
\label{llsum} \end{equation}
% Absolute value needed for correct u channel singularity

The individual terms in (\ref{llsum}) are given by
\begin{eqnarray}
{\cal L}_{i} &=& {2e^8 \over s{\widehat s} D(p, p')}
 \left[ f(p, p') \left|{\overline{\cal M}}_B({\widehat s}, t'_i)\right|^2
      + f(p', p) \left|{\overline{\cal M}}_B({\widehat s}, t_i)\right|^2
\right], \nonumber\\
{\cal L}_{f} &=& {2e^8 \over s{\widehat s} D(q, q')}
 \left[ f(-q', -q) \left|{\overline{\cal M}}_B(s, t_f)\right|^2
      + f(-q, -q') \left|{\overline{\cal M}}_B(s, t'_f)\right|^2
\right], \nonumber\\
{\cal L}_{p} &=& {2e^8 \over t t' D(p', q')}
 \left[ f(-q', p') \left|{\overline{\cal M}}_B(s_p, t)\right|^2
      + f(p', -q') \left|{\overline{\cal M}}_B({\widehat s}_p, t)\right|^2
\right], \nonumber\\
{\cal L}_{e} &=& {2e^8 \over t t' D(p, q)}
 \left[ f(p, -q) \left|{\overline{\cal M}}_B({\widehat s}_e, t')\right|^2
      + f(-q, p) \left|{\overline{\cal M}}_B(s_e, t')\right|^2
\right], \nonumber\\
{\cal L}_{m} &=& {2e^8 \over u u' D(p, q')}
 \left[ f(-q', p) \left|{\overline{\cal M}}_B(s_m, t_m)\right|^2
      + f(p, -q') \left|{\overline{\cal M}}_B({\widehat s}_m, t'_m)\right|^2
\right], \nonumber\\
{\cal L}_{m'} &=& {2e^8 \over u u' D(p', q)}
 \left[ f(p', -q) \left|{\overline{\cal M}}_B({\widehat s}_{m'}, t_{m'})
\right|^2
      + f(-q, p') \left|{\overline{\cal M}}_B(s_{m'}, t'_{m'})\right|^2
\right] \nonumber\\
 & & \label{llterms}
\end{eqnarray}
where the notation is defined as follows. The basic $s, t, u$ invariants
are defined by (\ref{tdef}), (\ref{sdef}) and
\begin{equation}
u = (p - q')^2, \qquad u' = (p' - q)^2.
\label{udef}\end{equation}
The pure Born amplitude with no photons is \({\cal M}_B(s_B, t_B)\),
where the effective Born parameters $s_B$, $t_B$ are as shown, with
\begin{eqnarray}
&t_i = \displaystyle{-{\widehat s} t\over t+u}, \qquad
t'_i = \displaystyle{-{\widehat s} t'\over t'+u'}, \quad
&t_f = \displaystyle{-st\over t+u'}, \qquad
t'_f = \displaystyle{-st'\over t'+u}, \qquad \label{Bornpar_a}\\
&s_p = \displaystyle{-st\over s+u'}, \qquad
{\widehat s}_p = \displaystyle{-{\widehat s} t\over{\widehat s}+u}, \quad
&s_e = \displaystyle{-st'\over s+u}, \qquad
{\widehat s}_e = \displaystyle{-{\widehat s} t'\over{\widehat s}+u'},
\qquad \label{Bornpar_b}\\
&s_m = \displaystyle{-su'\over s+t}, \qquad
{\widehat s}_m = \displaystyle{-{\widehat s} u'\over{\widehat s}+t'}, \quad
& t_m = \displaystyle{-tu'\over s+t}, \qquad
t'_m = \displaystyle{-t'u'\over {\widehat s}+t'}, \qquad \label{Bornpar_c}\\
&s_{m'} = \displaystyle{-su\over s+t'}, \qquad
{\widehat s}_{m'} = \displaystyle{-{\widehat s} u\over{\widehat s}+t}, \quad
& t_{m'} = \displaystyle{-tu\over{\widehat s}+t}, \qquad
t'_{m'} = \displaystyle{-t'u\over s+t'}. \qquad \label{Bornpar_d}
\end{eqnarray}
The denominator factor $D(p_1, p_2)$ is defined to be
\begin{equation}
D(p_1, p_2) = (p_1\cdot k_1)(p_1\cdot k_2)(p_2\cdot k_1)(p_2\cdot k_2)
              (p_1\cdot p_2)^{-4},
\label{sfactor}\end{equation}
and if the indicies \((i,j) = (1,2)\) or \((2,1)\) are chosen so that
\(p_1\cdot k_i + p_2\cdot k_i\) $\ge$ \(p_1\cdot k_j +
p_2\cdot k_j\), the form factors $f(p_1, p_2)$ are given by
\samepage{\begin{eqnarray}
f(p_1, p_2) &=& Y\left( {p_1\cdot k_i\over p_1\cdot p_2},
            \quad{p_1\cdot k_j\over p_1\cdot(p_2 - k_i)},
            \quad{p_2\cdot k_j\over p_2\cdot(p_1 - k_i)} \right) \nonumber\\
            &+& Y\left( {p_1\cdot k_i\over p_1\cdot(p_2 - k_j)},
            \quad{p_1\cdot k_j\over p_1\cdot p_2},
            \quad{p_2\cdot k_j\over p_1\cdot p_2} \right),
\label{ffactor}
\end{eqnarray}}
where \(Y(x,u,v) = (1 - x)^2 \left[ (1 - u)^2 + (1 - v)^2 \right]\).

We should note that, unlike the original LL expression in Ref.\ \cite{yfs2},
the result (\ref{lldcs}) does not control the next-to-leading-log corrections.
We will now discuss the comparison of exact and leading log
differential cross sections (\ref{dcs}) and (\ref{lldcs}), respectively,
in the SLC/LEP luminosity regime.

We have a made a detailed comparison of the exact and leading log
versions of the cross section (\ref{dcs})
in the luminosity regime. We illustrate these comparisons in Figs.\ 1
and 2. In Fig.\ 1, the two photons are emitted near the incoming
$e^+$ line in direction, with angles as given in the figure whereas
in Fig.\ 2, one photon is emitted along the direction of each
incoming charged particle, with the respective angles as given in the
figure. The final particles in the events associated with Figs.\ 1 and 2
are all taken to fall within the typical LEP-type trigger region as
described in Ref.\ \cite{theor_prec} and as we indicate in the figures.
%%## here I have refrased some important statements S.J.
What we see from Figs.\ 1 and 2, cases 1 and 2,
is that the LL result is always within
$20\%$ of the exact result in the relevant region of the phase space.
We have verified that this is true throughout the wide range of the
kinematic variables in the respective region of phase space.
This indicates that the error estimate for the
next-to-leading (NLL) ${\cal O}(\alpha^2)$
double bremsstrahlung effects not included in BHLUMI 2.01 in
Ref.\ \cite{theor_prec} is rather conservative.
The case 3 in Figs. 1 and 2 is relevant for next-to-next-to-leading (NNLL)
${\cal O}(\alpha^2)$ corrections which are at the level of $10^{-4}$ of the
integrated cross section.  The present calculation opens a path to
implementation of the NLL ${\cal O}(\alpha^2)$ double hard bremmsstrahlung
effect into BHLUMI in order to move it closer to the desired $.05\%$ precision
tag needed to support the $.15\%$ experimental errors expected in the
near term  high precision $Z^0$ physics program at LEP. This
implementation will appear elsewhere\cite{jwwws}.

The last curves in the figures show the size of the big logarithm
$L=\ln(|t|/m_e^2)-1$ in the respective regions of the phase space.
We see that $L$ varies significantly and, hence, that giving it
the proper argument $|t|$ instead of $s$ in the low angle regime suppresses
possible next-to-leading-log corrections. This then is consistent with the
implementation of YFS exponentiation in BHLUMI ,where
two of us (S. J. and B. F. L. W.) have also found that the proper
argument for $L$ in the respective radiation probablity is $|t|$.

We should also point out that our exact $2\gamma$ emission results
in low angle Bhabha scattering are directly relevant to the QED
expectations for high mass $2\gamma$ states in wide angle Bhabha
scattering at $Z^0$ energies. Thus, recent interest\cite{wide_angle}
in such events warrants a detailed assessment of such phenomena using
our results in this paper. This assessment will appear elsewhere\cite{jwwws}.
Our main objective in this paper is to present the exact results
for $e^+e^-\rightarrow e^+e^-+2\gamma$ at order $\alpha^2$ and to assess
the error estimate of the respective missing part of the order $\alpha^2$
bremsstrahlung effect in BHLUMI as it is given in Ref.\ \cite{theor_prec}.

\section{Conclusions}\label{CONCLUSIONS}    % section 5

In this paper we have used the methods of the CALKUL collaboration to
compute the important process
of two photon bremsstrahlung in Bhabha scattering in the $Z^0$ resonance
region at ${\cal O}(\alpha^2)$. We have compared our exact results for
the corresponding differential cross section with leading log (LL)
expectations as a check of the results in Ref.\ \cite{theor_prec} on the
total precision of the Monte Carlo program BHLUMI2.01\cite{bhlumi}.

Specifically, we have presented new formulas for the process
$e^+e^-\rightarrow e^+e^-+2\gamma$ at ${\cal O}(\alpha^2$) in the $Z^0$
resonance region. We have compared our exact results with the LL
expectations for the respective differential cross section and we find
that the two differential cross sections are within $20\%$ of each
other throughout the relevant region of the phase space. This indicates
that the estimate in Ref.\ \cite{theor_prec} for the contribution of
the missing NLL ${\cal O}(\alpha^2)$ bremsstrahlung in BHLUMI2.01\cite{bhlumi}
to its total precision of $.25\%$ is rather conservative.
The implementation of  this missing part of the ${\cal O}(\alpha^2)$
bremsstrahlung into BHLUMI and will be report elsewhere
\cite{jwwws}.

We emphasize that our new formulas are valid both at low and wide
electron, positron scattering angles as defined by the typical LEP/SLC
trigger in the luminosity regime, {\em i.e.}, low refers to scattering angles
in the luminosity regime and wide refers to such angles larger than the
trigger angles. This means that our results have an immediate application
to the recent discussion of large two-photon mass wide angle lepton pair
production events in Ref.\ \cite{wide_angle}. Such applications will appear
elsewhere\cite{jwwws}.

In summary, the exact results which we have presented for the two
photon bremsstrahlung process in Bhabha scattering in the $Z^0$ resonance
region allow us to make an important step in reducing the error
on the teoretical prediction for the LEP/SLC luminosity process to
below $.1\%$ regime. We look forward to the attendant refinement
in the precision of the respective Standard Model tests in $Z^0$ physics.

%\appendix            % Further sections are appendices (if any):
%
%\section{}           \label{DETAILS}        % appendix A
%
% ... more appendices, if any

\section*{Acknowledgments}
We acknowledge useful interactions and discussions with Dr.\ S. Lomatch
in the early phases of this work. Two of us (S. J. and B. F. L. W.)
are grateful to Profs. M. Breidenbach, J. Dorfan, J. Ellis, G. Feldman
and C. Prescott for the opportunity to participate in the MKII-SLD SLC
and CERN LEP Physics Working Groups wherein the early stages of this
work were generated. These same two authors are also grateful to
Prof.\ F. Gilman of the SSCL for its kind hospitality
while part of this work was completed.

\newpage
\section*{Figure captions}

\par\vspace{.5cm}\noindent
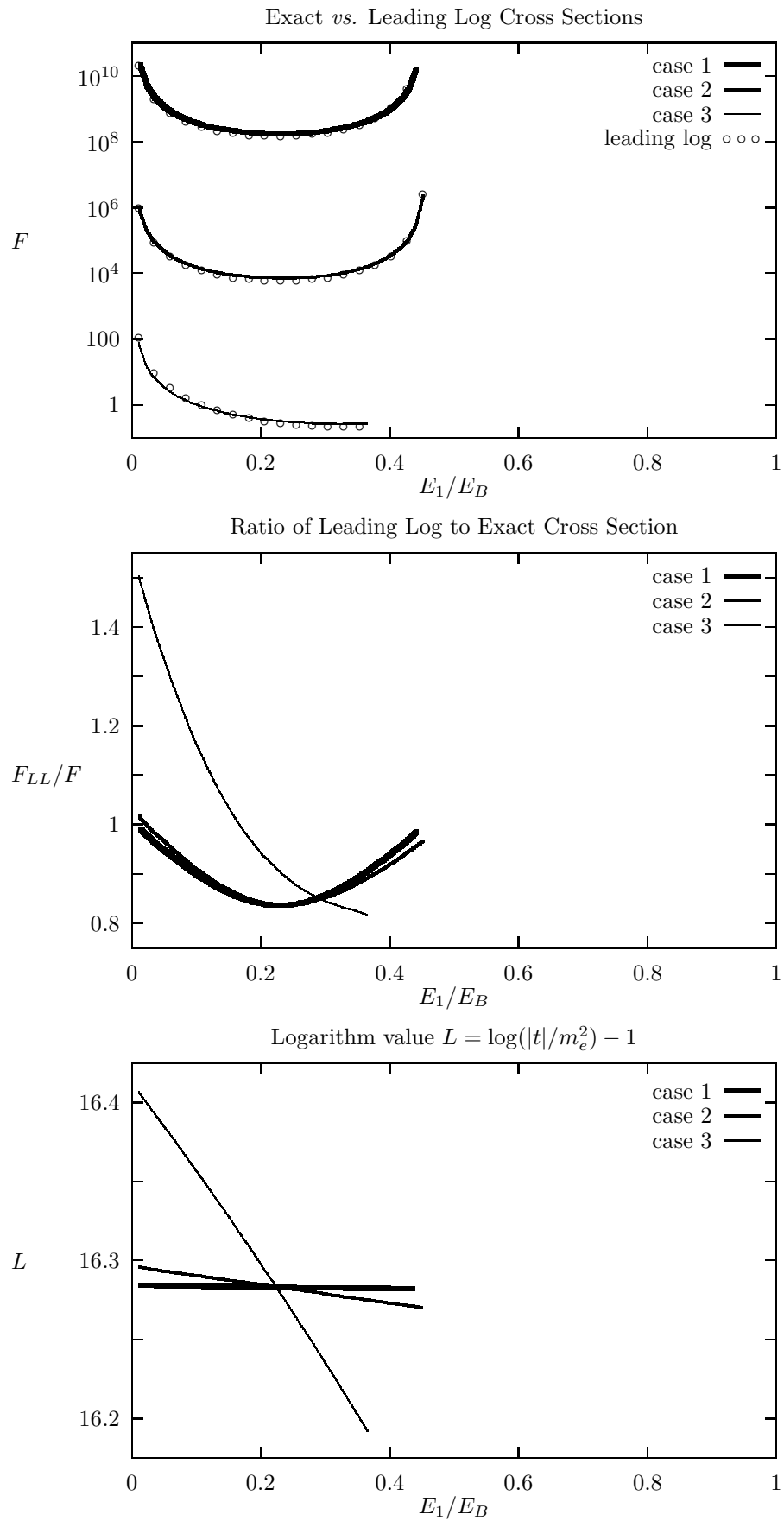
Fig.\ 1. Comparison of LL and exact results for low angle
Bhabha scattering in the SLC/LEP luminosity regime. The functions
$F$ are the normalized differential cross-sections defined in
Ref.\ \cite{s-channel}.
%%##%%##
We parametrize 8-dimensional $e^+e^-\gamma\gamma$ phase space in terms
of seven angles and the energy of one photon $E_1$.
We plot dependence on photon energy $E_1$ for fixed values of
angles which are listed in the following table.
Here, in the table, the subcripts 1 and 2 refer to the
photons and where the photon angles are relative
to the incoming positron direction so that, in this figure, both
photons are emitted along the incoming positron direction:
\begin{center}
Fig. 1 Kinematic Variables \\
\hfill\\
\begin{tabular}[h]{|c||cc|cc|cc|c|}\hline
case & $\theta_1$ & $\phi_1$ & $\theta_2$ & $\phi_2$ &
$\theta_{e^+}$ & $\phi_{e^+}$ & $\theta_{e^-}$ \\ \hline
  1&$    6.4\times 10^{-3\ \circ}$&$  0^\circ$&$   6.4\times 10^{-3\ \circ}$&$
         180^\circ$&$    4.9^\circ$&$  0^\circ$&$   2.7^\circ$\\
  2&$    8\times 10^{-2\ \circ}$&$  0^\circ$&$   8\times 10^{-2\ \circ}$&$
         180^\circ$&$    4.9^\circ$&$  0^\circ$&$   2.7^\circ$\\
  3&$    1^\circ$&$  0^\circ$&$    1^\circ$&$   180^\circ$&$
         4.9^\circ$&$  0^\circ$&$   2.7^\circ$\\
\hline
\end{tabular} \\
\hfill\\
\end{center}
Note that the energy of the outgoing charged particles is required
to remain above $0.5$ of the beam energy $E_B$. For the
three cases 1, 2, 3 of kinematics shown in the table, the plots show
respectively the individual exact and LL cross sections, their
ratio.
The cases 1 and 2 are relevant to NLL correction while case 3 covers
NNLL region of the phase space.
The size of the dominant big logarithm $\ln(|t|/{m_e}^2)-1$
which determines the probability to radiate in the process is also plotted.

\par\vspace{1cm}\noindent
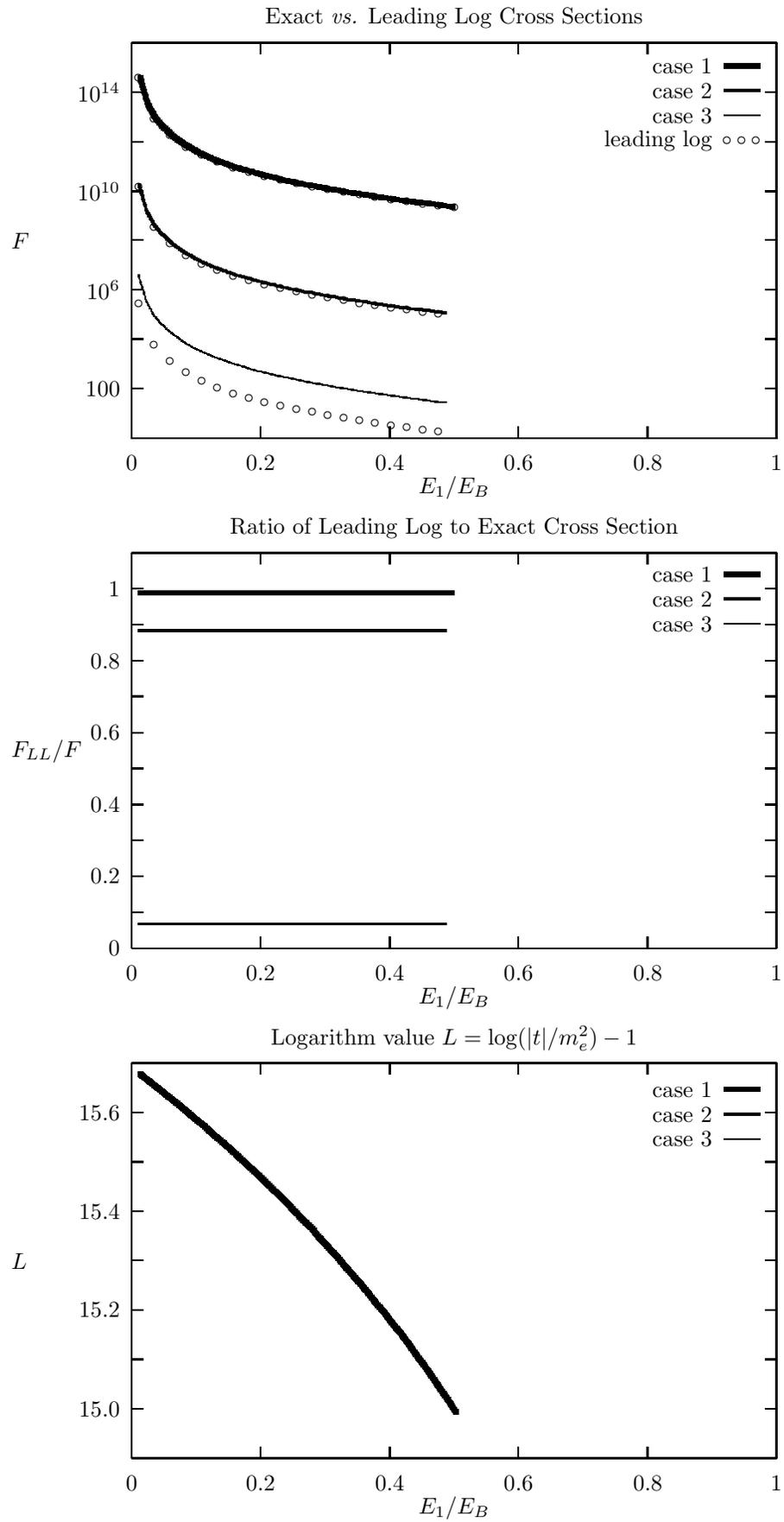
Fig.\ 2. The same plots as those given in Fig. 1 with photon 1's
spherical angles measured relative to the incoming positron direction
and with photon 2's angles measured relative to the incoming
electron direction. Thus, the two photons are emitted in generally
opposite directions. The specific kinematic input is summarized in
the following table:
\begin{center}
Fig. 2 Kinematic Variables \\
\hfill\\
\begin{tabular}[h]{|c||cc|cc|cc|c|}\hline
case & $\theta_1$ & $\phi_1$ & $\theta_2$ & $\phi_2$ &
$\theta_{e^+}$ & $\phi_{e^+}$ & $\theta_{e^-}$ \\ \hline
  1&$  6.4\times 10^{-3\ \circ}$&$  0^\circ$&$    6.4\times 10^{-3\ \circ}$&$
         180^\circ$&$    2.7^\circ$&$  0^\circ$&$  2.7^\circ$\\
  2&$  8\times 10^{-2\ \circ}$&$  0^\circ$&$    8\times 10^{-2\ \circ}$&$
         180^\circ$&$    2.7^\circ$&$  0^\circ$&$  2.7^\circ$\\
  3&$  1^\circ$&$  0^\circ$&$    1^\circ$&$    180^\circ$&$
         2.7^\circ$&$  0^\circ$&$  2.7^\circ$\\
\hline
\end{tabular} \\
\hfill\\
\end{center}
The cut on the charged particles' outgoing energies is the
same as in Fig.\ 1.

\vfill\eject
% \end{document}
%%%%%%%%%%%%%%%%%%%%%%%%%%%%%%%%%%%%%%%%%%%%%%%%%%%%%%%%%%%%%%%%%%%%%%
% The figures follow. If this file is too long to TeX, cut here, and
% uncomment the \end{document} statement above and the following line:
% \documentstyle{article} \begin{document}

\font\tenrm=cmr10
\pagestyle{empty}
\begin{figure}
\centering
\vspace{-3cm}
% GNUPLOT: LaTeX picture
\setlength{\unitlength}{0.240900pt}
%
\vskip 10pt
Figure 2
\end{figure}
\clearpage
\end{document}